\documentstyle[aps,epsf]{revtex}
\begin{document}

\draft

\title{Dynamical Chiral Symmetry Breaking on a Brane in Reduced QED}

\author{E.V. Gorbar$^{*}$}
\address{ Departamento de Fisica, Universidade Federal de Juiz de Fora,
36036-330, MG, Brazil} 
\author{V.P. Gusynin$^{*}$} 
\address{ Department of Physics, Nagoya University, Nagoya 464-8602,
Japan}
\author{V.A.~Miransky$^{*}$}
\address{Department of Applied Mathematics, University of Western
Ontario,\\ London, Ontario N6A 5B7, Canada}

\date{\today}
\maketitle
\begin{abstract} 
Reduced gauge theories are theories in which while gauge fields
propagate in a bulk, fermion fields are localized on a brane.
We study dynamical chiral symmetry breaking on a 2-brane  and
a 1-brane in reduced $\mbox{QED}_{3+1}$, and on a 1-brane in
reduced $\mbox{QED}_{2+1}$. Since, unlike higher dimensional
gauge theories, $\mbox{QED}_{3+1}$ and  $\mbox{QED}_{2+1}$
are well defined, their reduced versions can serve as a
laboratory for studying dynamics in a higher dimensional
brane world. The analysis of the Schwinger-Dyson (SD) equations
in these theories reveals rich and quite nontrivial
dynamics in which the conformal symmetry and its breakdown
play a crucial role.
Explicit solutions of the SD equations in the
near-critical regime are obtained and the character of the
corresponding phase transition is described.   
\end{abstract}

\section{Introduction}

Dynamics in a brane world has recently attracted considerable
interest. In most cases, it has been studied in higher dimensional
theories (for a recent review, see Ref. \cite{Rubakov}).
The aim of this paper is to consider dynamics on a brane
not in higher dimensions but in a 3+1 and 2+1 dimensional world.
More precisely, we study dynamical chiral symmetry breaking in
the reduced $\mbox{QED}_{3+1}$ and $\mbox{QED}_{2+1}$.
The term "reduced" implies here that while massless gauge fields 
propagate in a (3+1 or 2+1 dimensional) bulk, fermion fields are 
localized on a brane. We will consider the cases of a 2-brane and 
a 1-brane in $\mbox{QED}_{3+1}$ and a 1-brane in $\mbox{QED}_{2+1}$. 
We would like also to emphasize that though we use the conventional 
term $\mbox{"QED"}$ for a $U(1)$ abelian gauge theory with fermions, 
we do not specify the origin of the gauge field: it is not necessary 
electromagnetic field.

Motivations for considering this type of models are rather
obvious. It is well known that relativistic field models can
serve as effective theories for the description of long 
wavelength excitations in condensed matter systems \cite{CM}.
The reduced $\mbox{QED}$ describes the situation when while fermions
are localized on a plane (say, on a $\mbox{Cu-O}$ plane in a 
high-$T_{c}$ superconductor) or on a string (polymer like systems), 
interactions between them are provided by a bulk gauge field. Besides 
that, reduced $\mbox{QED}_{3+1}$ can be relevant for the dynamics of 
cosmological strings \cite{strings}. At last, reduced $\mbox{QED}$ 
has been recently considered in higher dimensions for the description 
of the mechanism of (quasi)localization of a photon field on a
3-brane \cite{DGS} .

Another, more practical, reason is using the reduced $\mbox{QED}_{3+1}$ 
and $\mbox{QED}_{2+1}$ as a laboratory for studying dynamical 
chiral symmetry breaking in the brane world. Unlike higher
dimensional gauge theories, $\mbox{QED}_{3+1}$ and $\mbox{QED}_{2+1}$
are well defined. While $\mbox{QED}_{3+1}$ is renormalizable and
therefore well defined in perturbative theory, $\mbox{QED}_{2+1}$  is
superrenormalizable and therefore asymptotically free.
 
As we will see, the dynamics of chiral symmetry breaking in
reduced QED is quite nontrivial. In particular, the conformal
symmetry (and its breakdown) plays a crucial role in the dynamics.

The paper is organized as follows. In Sec. \ref{features} general
features of reduced $\mbox{QED}$ are described. In Sec. 
\ref{42} we study dynamical chiral symmetry breaking on a 2-brane
in reduced $\mbox{QED}_{3+1}$. In Sec. \ref{41} chiral symmetry
breaking on a 1-brane in reduced $\mbox{QED}_{3+1}$ is
considered. In particular, we discuss subtleties connected
with spontaneous breakdown of continuous symmetries on a
1-brane. In Sec. \ref{31} chiral symmetry breaking in the
reduced  $\mbox{QED}_{2+1}$ with a 1-brane is studied. In
Sec. \ref{conclusion} we summarize our results. 
An analysis of the Schwinger-Dyson equation in 
the reduced  $\mbox{QED}_{2+1}$ with a 1-brane is done
in Appendix A.
\section{Reduced QED: general features}
\label{features}

In this section, general features of the reduced QED will be
described.
The $\mbox{QED}_{(D-1)+1}$ action in Euclidean space reads 
($X=(x_D, x_1,..., x_{D-1}$))
\begin{equation}
S = \int d^{D}X (\frac{1}{4e^2} F_{ab}^2 + A_{a}J^{a} -
\frac{1}{2e^{2}\xi} (\partial_{a}A^{a})^2),
\label{in_action}
\end{equation}
where $\xi$ is a gauge parameter and $J^{a}$ is a fermion current. We
will consider the chiral limit (no fermion bare mass term) and, for
convenience, consistently omit the kinetic term for
fermions in the action, restoring it only when it is necessary. 
Integrating over $A_{a}$, we get
\begin{equation}
S = \frac{1}{2} \int d^{D}X d^{D}Y
J^{a}(X)\tilde{D}_{ab}^{(0)}(X-Y)J^{b}(Y),
\label{action_source}
\end{equation}
where 
\begin{equation}
\tilde{D}_{ab}^{(0)} = e^2 \int \frac{d^DK}{(2\pi)^D} 
\exp\left({iK(X-Y)}\right) 
\left(\delta_{ab} -
(1 - {\xi})\frac{K_{a}K_{b}}{K^2}\right)\frac{1}{K^2}
\label{D-propagator}
\end{equation}
with $K= (k_D, k_1, ..., k_{D-1})$.
In reduced $\mbox{QED}$, with a d-brane, we assume that the fermion
current has the following form:
\begin{eqnarray*}
J^{a}(X)=0 \,\,\, \mbox{for} \,\,\, a = d+1,d+2,...,D-1,
\end{eqnarray*}
\begin{equation}
J^{a}(X)=j^{a}(x_D,x_1,...,x_d)\delta^{D-d-1}(\bar{x})
\,\,\, \mbox{for}
\,\,\,
a = D,1,...,d,
\label{current}
\end{equation}
where $\bar{x}\equiv (x_{d+1},...,x_{D-1})$.
Integrating over $\bar{x}$ and $\bar{y}$ in Eq.(\ref{action_source}),
we obtain the reduced d+1 dimensional action
\begin{equation}
\tilde{S}_{[Dd]eff}= \frac{1}{2} \int
d^{d+1}xd^{d+1}y j^{\mu}(x)D_{[Dd]\mu\nu}^{(0)}(x - y)j^{\nu}(y),
\label{redaction}
\end{equation}
where
\begin{equation}
{D}_{[Dd]\mu\nu}^{(0)}(x - y) = e^2 \int\frac{d^{d+1}k
d^{D-d-1}\bar{k}}
{(2\pi)^D} \exp\left({ik(x-y)}\right)                     
\left(\delta_{\mu\nu} -(1-
\xi)\frac{k_\mu k_\nu}{\bar{k}^2 + k^2}\right)
\frac{1}{\bar{k}^2 + k^2}
\label{barepropagator}
\end{equation}
with $\mu,\nu=D,1,...,d$ (the notation for momenta we use here is 
self-explanatory).

As it will be shown in the next sections, after integrating over
$\bar{k}$ momenta, the effective action can
be rewritten in the following general form:
\begin{equation}
S_{[Dd]eff} = \int d^{d+1}x \left[\frac{1}{4e^2}F_{\mu\nu}
I(-\partial^2)F^{\mu\nu}
+ A_{\mu}j^{\mu} + gauge\quad{term} \right],
\label{effaction}
\end{equation}
where $\partial^2$ is the Laplacian in d+1 dimensions and
$I(-\partial^2)$ is a non-local (i.e. integral) operator.

The following properties of the action (\ref{effaction}) are
noticeable:

a) The interacting term $A_{\mu}j^{\mu}$ is conformally
invariant for all $D$ and $d$. This point will be
important for the dynamics of chiral symmetry breaking in
reduced $\mbox{QED}$.

b) When $d=D-2$, the kinetic term in expression (\ref{effaction})
is finite. However, when $d<D-2$, there are ultraviolet divergences
in it. The reason for that is simple. Because of a delta
function in the fermion current (\ref{current}), 
integrating over $\bar{x}=(x_{d+1},...,x_{D-1})$ encounters a classical 
self energy of a point like particle in $D-d-1$ dimensions. It is
finite in the one dimensional case ($d=D-2$) and divergent otherwise. 
Therefore, when $d<D-2$, one should regularize the delta function, i.e.
introduce a finite thickness for a d-brane. This is an additional
source of the breakdown of conformal symmetry. Notice that in the
reduced $\mbox{QED}_{3+1}$ with $d=2$, the kinetic term is both finite 
and conformally invariant.    

c) Effective action (\ref{effaction}) describes fermion fields and 
a projection of the gauge field on a brane. Since gauge bosons can escape 
from the brane to the bulk, the unitarity does not fulfill in the
brane dynamics. In the next section, we will discuss explicit
manifestations of this feature of reduced $\mbox{QED}$. 
 
\section{Reduced $\mbox{QED}_{3+1}$ with a 2-brane}
\label{42}

In this section we will study spontaneous chiral symmetry 
breaking in the reduced $\mbox{QED}_{3+1}$ with a 2-brane, i.e.
with $D=4$, $d=2$ and $\bar{k}=k_{3}$. 
Integrating over $k_3$ in expression (\ref{barepropagator}),
we obtain the bare gauge field propagator of an effective 2+1 
dimensional theory on a 2-brane:
\begin{equation}
D_{[42]\mu\nu}^{(0)}(x - y) = \frac{e^2}{2}\int\frac{d^3k}{(2\pi)^3} 
\exp\left({ik(x-y)}\right)
\left(\delta_{\mu\nu} -(1- \xi)\frac{k_\mu k_\nu}{k^2}
\right)\frac{1}{\sqrt{k^2}},
\label{42barepropagator}
\end{equation}
where $\mu,\nu=4,1,2$ and, for convenience, we made the substitution
$\xi\to{2\xi-1}$, i.e. $(1-\xi)\to 2(1-\xi)$. 
Introducing a 2+1 vector field $A_{\mu}(x)$,
the effective action (\ref{redaction}) can be rewritten
in the following form (compare with Eq. (\ref{effaction})): 
\begin{equation}
S_{[42]eff} = \int d^3x \left[\frac{1}{2e^2}F_{\mu\nu}
\frac{1}{\sqrt{-\partial^2}}F^{\mu\nu} + A_{\mu}j^{\mu} +
\frac{1}{e^2\xi}\partial_{\mu}A^{\mu}
\frac{1}{\sqrt{-\partial^2}}\partial_{\nu}A^{\nu} \right].
\label{42effaction}
\end{equation}
One should add the kinetic term of fermions on a 2-brane
to this action:
\begin{equation}
S_{kin}= \int d^3x\bar{\psi}\left(i\gamma^{\mu}\partial_{\mu}\right)\psi.
\label{3fermions}
\end{equation}
As is well known, there are two (two dimensional) inequivalent
representations of the Clifford algebra in 2+1 dimensions.
Following Refs. \cite{JT,P,ABKW}, we will consider
four component fermion fields which contain these two
inequivalent representations. In this case, there
exists a fermion mass term preserving parity. If there are
$N_f$ fermion flavors, the symmetry of the action is
$U(2N_{f})$ \cite{P,ABKW}. The dynamical generation of a
fermion mass leads to spontaneous breakdown of this symmetry
down to $U(N_{f}) \times U(N_{f})$.

A remarkable feature of the action (\ref{42effaction})
is that it is conformal invariant. Since the initial $\mbox{QED}$
theory is renormalizable, one should expect that this feature plays an 
important role in the dynamics.
Our aim is to describe spontaneous chiral symmetry breaking in the
theory with this effective action. Since there is no dimensional
parameters in the action (\ref{42effaction}), a fermion dynamical
mass $m_d$ can be induced only through the mechanism of
dynamical transmutation. In our case, it means that one should
intoduce an ultraviolet cutoff $\Lambda$, thus breaking the
conformal symmetry. Then the dynamical mass, if it arises at
all, will be proportional to $\Lambda$. We will be especially
interested in the near-critical regime of the dynamics, when
$m_d\ll \Lambda$. 

The dynamical chiral symmetry breaking in this model is
a highly nontrivial problem,
and our strategy for solving it (at least approximately)
will be to find a framework in which the improved ladder
(rainbow) approximation would be reliable. We recall that while in
the ladder (rainbow) approximation there is only a Schwinger-Dyson (SD)
equation for the fermion propagator (with both the vertex and the gauge
field propagator being bare), in the improved ladder (rainbow)
approximation there are two SD equations (with a bare vertex),
both for the fermion and for the gauge field propagators.

The SD equation for the fermion propagator in Minkowski
space in the improved ladder approximation has the following form:
\begin{equation}
G^{-1}(p) = G^{(0)^{-1}}(p)+i\int\frac{d^3q}{(2\pi)^3}\gamma^{\mu}G(q)
\gamma^{\nu}D_{[42]\mu\nu}(p-q),
\label{42SDequation}
\end{equation}
where $G^{(0)}(p)$ is the bare fermion propagator and 
\begin{equation}
D_{[42]\mu\nu} = \left( g_{\mu\nu} - (1 -\xi(k^2))
\frac{k_{\mu}k_{\nu}}{k^2} \right)D(k^2)
\label{full_ph_prop}
\end{equation}
is the full gauge field propagator for which there is its own SD
equation (with a bare vertex in this approximation). Here we use
a general non-local gauge with $\xi(k^2)$ being a function of
$k^2$ (a need for considering such gauges will soon become clear: 
see Eq. (\ref{gauge}) below).
The bare gauge field propagator is now (compare with
Eq. (\ref{42barepropagator}))
\begin{equation}
D_{[42]\mu\nu}^{(0)} = \left( g_{\mu\nu} - (1 -
\xi(k^2))\frac{k_{\mu}k_{\nu}}{k^2} \right) \frac{e^2}{2\sqrt{-k^2}}
\end{equation}
and the full gauge field propagator is related to the vacuum polarization
tensor $\Pi_{\mu\nu}(k)$:  
\begin{equation}
D^{-1}_{[42]\mu\nu}(k)
=D^{(0)^{-1}}_{[42]\mu\nu}(k)+\Pi_{\mu\nu}(k),\quad
\Pi_{\mu\nu}(k)=\left(g_{\mu\nu}-\frac{k_\mu k_\nu}{k^2}\right)\Pi(k^2).
\end{equation}
The structure of the propagator $G(p^2)$ is
$G(p^2) = (A(p^2)\hat{p}-B(p^2))^{-1}$, $\hat{p} \equiv
\gamma^{\mu}p_{\mu}$, 
and
from Eqs. (\ref{42SDequation}) and (\ref{full_ph_prop}) we obtain  
the following equations for $A(p^2)$ and $B(p^2)$ in the Euclidean space
($p_{0}=ip_{4}$):
\begin{eqnarray}
A(p^2) &=& 1 + \frac{1}{p^2}\int\frac{d^3q}{(2\pi)^3}
\frac{A(q^2)}{q^2A^2(q^2) + B^2(q^2)}D[(p-q)^2] 
\nonumber \\
&\times& \left( pq + (1- \xi((p-q)^2))(pq - \frac{2(p^2q^2 -
(pq)^2)}{(p-q)^2})\right),
\label{Aequation}
\end{eqnarray}
\begin{equation}
B(p^2) = \int\frac{d^3q}{(2\pi)^3}\frac{B(q^2)}{q^2A^2(q^2) + B^2(q^2)} 
D[(p-q)^2]\left( 2 + \xi((p-q)^2) \right).
\label{Bequation}
\end{equation}
Notice that the function $D(k^2)$ is expressed through the
vacuum polarization function $\Pi(k^2)$ as
\begin{equation}
D(k^2) = \frac{1}{\frac{2k}{e^2} + \Pi(k^2)}.  
\label{invcharge}
\end{equation}
The full gauge field propagator (\ref{full_ph_prop}) 
satisfies its own SD equation and
therefore is in principle a complicated functional of the
functions $A(p^2)$ and  $B(p^2)$. Fortunately, in the present case
the situation can be considerably simplified. First of all, as
we will see below, one can choose a gauge in which the function
$A(p^2)$ is identically equal to 1, and we will use such a 
gauge. Second, it will be shown that, in the near-critical
regime (when $m_d\ll \Lambda$), the fermion
dynamical mass, defined as $m_d=B(m^{2}_{d})$,
is mainly induced in the kinematic region with
$m_d^2\ll k^2$. In that  region, fermions can be treated as
massless, and, if $A(p^2)=1$ , the polarization
function is given by the one loop expression with the fermion
propagators of free, 2+1 dimensional, massless fermions.

For completeness and convenience, however, we will use
the one loop expression for $\Pi(k^2)$ taking free fermions with
the mass $m = m_d$. On a 2-brane, i.e. in 2+1 dimensions, it
is: 
\begin{equation}
\Pi(k^2) = \frac{N_f}{4\pi} [2m_d + \frac{k^2 - 4m_d^2}{k}
\mbox{arctan}\frac{k}{2m_d} ].
\label{polarization}
\end{equation}
Notice that
\begin{equation}
\Pi(k^2) \to \frac{N_{f}k}{8} 
\label{uvasymptotics}
\end{equation}
for $k \gg m_d$, and
\begin{equation}
\Pi(k^2) \to \frac{N_{f}k^{2}}{6\pi m_d} 
\label{irasymptotics}
\end{equation}
for $k \ll m_d$.

When can the improved ladder approximation be reliable? The simplest
case would be of course the dynamics with a small coupling constant
$\alpha = e^2/{4\pi}$ (notice that $\alpha$ is a bare coupling 
constant here). In that case, even the ladder (rainbow)
approximation would be good enough. Unfortunately, as it will be
shown below, for small $\alpha$ there is no solution with spontaneous chiral
symmetry breaking in the reduced $\mbox{QED}_{3+1}$ with a 2-brane.
Therefore one should try something else. 

Our initial observation is
that the structure of SD equations (\ref{Aequation})
and (\ref{Bequation}) is similar to that in usual,
non-reduced, $\mbox{QED}_{2+1}$. It had been recognized long ago that
the $1/N_{f}$ expansion can be useful in that theory
\cite{P,ABKW,ANW,nash}. Though being very   
nontrivial, the $1/N_{f}$ expansion is helpful in putting under control  
of nonperturbative dynamics. The crucial point is the selection of a
``right"
gauge in the leading order in $1/N_{f}$, in which the improved ladder
approximation would be reliable \cite{QED3}. In particular, 
appropriate Ward identities have to be satisfied in that gauge. In
other gauges, the results can be found by gauge-transforming Green's
functions from the ``right" gauge
to those gauges. Such a transformation in general changes the initial 
improved ladder approximation to another one, though the gauge
invariant quantities remain of course the same. 
We will adopt this strategy for the present problem and, first
of all, check the Ward identity for the vertex. Since in this
approximation, by definition, the vertex is bare, the function
$A(p^2)$ has to be equal one. It is known \cite{gauge}, that
for the full photon propagator (\ref{full_ph_prop}), and in
arbitrary $d$ space dimensions,
this function is identically equal to 1 if one
uses a non-local (in general) gauge with the following gauge function
$\xi(k^2)$:
\begin{equation}
\xi(z) = d - \frac{d(d-1)}{z^{d}D(z)}\int_0^z dt t^{d-1}D(t).
\label{gauge}
\end{equation}

We will see that, in the near-critical regime,
the momentum region mostly responsible for the mass generation is $k \gg
m_d$. As it follows from Eq. (\ref{uvasymptotics}), in that region
$\Pi(k^2) = \frac{N_{f}k}{8}$, i.e. the   
function $D(k^2)$
(\ref{invcharge}) is proportional to $k^{-1}$.
For such a function $D(k^2)$ and $d=2$, one gets $\xi(k^2) = 2/3$
(the so called Nash gauge \cite{nash}), and the gap equation takes the
form
\begin{equation}
B(p^2) = 4\pi^{2}\lambda \int \frac{d^3q}{(2\pi)^3} \frac{B(q^2)}
{q^2 + B^2(q^2)} \frac{1}{\sqrt{(p-q)^2}}, \quad \lambda = \frac{e^2}
{3\pi^2(1 + \frac{N_f e^2}{16})}.
\label{SD}
\end{equation}
The validity of the Ward identity is a necessary but not
of course sufficient condition for the reliability of the
improved ladder approximation. The crucial point for that
is a justification of the use of a bare vertex. This approximation
for the vertex can be justified in the leading order of the
$1/N_{f}$ expansion \cite{P,ABKW,ANW,nash,QED3}.
 
Integrating over angles in Eq. (\ref{SD}), we obtain
\begin{equation}
B(p^2) = \lambda \int_{0}^{\Lambda^2} \frac{dq^2 \sqrt{q^2} B(q^2)}
{q^2 + B^2(q^2)} \frac{\sqrt{2}}{\sqrt{p^2 + q^2 +|p^2-q^2|}}.
\label{1SD}
\end{equation}
Here the ultraviolet cutoff $\Lambda$ was introduced.

Notice that in the momentum region $q^2 \gg
m_d^{2}\equiv B^2(m_d^{2})$, the term $B^{2}(q^2)$ in the 
denominator of the integrand of expression (\ref{1SD}) is
irrelevant. The only role of this term is to provide a cutoff in
the infrared region. Therefore one can drop this term, introducing
an explicit infrared cutoff in the integral. Then we obtain the following 
equation ($x=p^2, y=q^2$):
\begin{equation}
B(x) = \lambda \int_{m^2_d}^{\Lambda^2} \frac{dy}{y^{1/2}} B(y)
[\frac{\theta(x-y)}{\sqrt{x}}
+\frac{\theta(y-x)}{\sqrt{y}}].
\label{2SD}
\end{equation} 
The transition from equation (\ref{1SD}) to equation (\ref{2SD})
corresponds to the so called bifurcation approximation (or method).
For the problem of dynamical symmetry breaking, this method was 
introduced in Ref. \cite{Atkinson} and since then has been widely
used in the literature (for a review see Ref. \cite{book}). This 
method is especially appropriate for the near-critical dynamics: the
closer the dynamics is to a critical (bifurcation) point, the smaller
the dynamical mass $m_d$, and therefore the term $B^2(q^2)$ in the
denominator of the integrand (\ref{1SD}), become. 

It is easy to check that the integral equation (\ref{2SD}) is
equivalent to the differential equation 
\begin{equation}
x^2B^{\prime\prime} + \frac{3x}{2}B^{\prime} + \frac{\lambda}{2}B = 0
\label{diffeq}
\end{equation}
with the following two boundary conditions:
\begin{equation}
B^{\prime}(m^2_d) = 0,
\label{ir}
\end{equation}
\begin{equation}
(2xB^{\prime} + B)|_{x=\Lambda^2} = 0.
\label{uv}
\end{equation}
A solution of Eq.(\ref{diffeq}) which satisfies the 
infrared boundary condition (\ref{ir}) is
\begin{equation}
B(x) = \frac{m_d^{3/2}}{x^{1/4}\sinh\delta} \sinh\left(\frac{\omega}{4}\log\frac{x}{m_d^2} +
\delta\right),
\label{solution}
\end{equation}
where $\omega = \sqrt{1 - 8\lambda}$ and $\delta = \frac{1}{2} \log
\frac{1+\omega}{1-\omega}$,
and here we also used the normalization condition
$B(m^{2}_{d}) = m_d$.
The ultraviolet boundary condition (\ref{uv}) yields the following
equation for 
the dynamical mass:
\begin{equation} 
\mbox{th}(\frac{\omega}{2} \log \frac{\Lambda}{m_d} + \delta) = -
\omega.
\label{uv1}
\end{equation}
Obviously, there is no solution $m_d \ll \Lambda$
for $\lambda < \lambda_{cr} = 1/8$. For supercritical values
of $\lambda$
($\lambda > \lambda_{cr}$), Eq.(\ref{uv1}) takes the form
\begin{equation} 
\mbox{tg}(\frac{\nu}{2} \log \frac{\Lambda}{m_d} + \mbox{arctg}\nu) =
-\nu,
\end{equation}
where $ \nu = \sqrt{8\lambda - 1}$. Therefore for small $\nu$
the mass is
\begin{equation}
m_d\simeq
\Lambda \exp[- \frac{2\pi}{\nu} + 4].
\label{mass}
\end{equation}
The critical line in the plane ($N_f, \,\, e^2$) is given by
\begin{equation}
e_{cr}^2 = \frac{16}{N_{max} - N_f},
\label{crline}
\end{equation}
where $N_{max} = \frac{128}{3\pi^2}$. Spontaneous chiral symmetry
breaking takes place for $e > e_{cr}$, and
the value $N_{max}$ defines
the upper limit for the number of fermion flavors $N_{f}$ for
which spontaneous chiral symmetry breaking is possible. When
$N_f \to N_{max}$, the critical value 
$e^{2}_{cr} \to \infty$.  

Let us now discuss self-consistency of the assumption that the region of
momenta $q \gg m_d$ is mostly responsible for the generation
of the mass in the near-critical regime ($m_d \ll \Lambda$) . 
The point is that in this
regime $\ln{\Lambda/m_d} \sim {2\pi}/\nu$ is large. On the other
hand, 
the behavior of the integrand on the right hand side of equation
(\ref{1SD}) is smooth as $q^2 \to 0$. 
The smooth
behavior of the integrand in the infrared region implies that the
region $0 \leq q \alt m_d$ is too small to generate the
large logarithm $\ln{\Lambda/m_d}$.
It (and therefore the essential singularity in
expression (\ref{mass})) is generated in the large region 
$m_d \ll q \ll \Lambda$. A variation of the kernel
in the infrared region
can at most change the overall coefficient in that expression.
This heuristic argument is supported by numerical studies of
integral equation (\ref{1SD}).    

The critical line (\ref{crline}) implies that it is a strong
coupling dynamics, with $e^2 > e^{2}_{cr}$, that provides
spontaneous chiral symmetry breaking on a brane. Indeed, the lowest
$e^{2}_{cr}$ corresponds to $N_{f}=1$ and it is
$e^{2}_{cr} \simeq 4.81$, i.e. $\alpha_{cr} \equiv
{e^{2}_{cr}}/{4\pi} \simeq 0.38$. 

This strong coupling dynamics is provided by (essentially)
conformal invariant interactions 
in the most important region of momenta
$m_d \ll q \ll \Lambda$. Indeed, up to the irrelevant $B^{2}(q^2)$
term in the denominator, the kernel of
integral equation (\ref{SD}) transforms as $K(p^2,q^2) \to
K(s^{2}p^2,s^{2}q^2)=s^{-3}K(p^2,q^2)$
under the scale transformation $p,q \to
sp,sq$. This, together with the transformation of the measure
$d^{3}q \to s^{3}d^{3}q$, implies that the interactions are
indeed (essentially) conformal invariant 
in that region.[In the integral equation (\ref{2SD}),
the conformal symmetry is broken only by
the dimensional boundary parameters $\Lambda$ and $m_d$ in
the integral.] This reflects the presence of long range, Coulomb
like, interactions which provide the essential singularity
in expression (\ref{mass}).

The critical line (\ref{crline}) corresponds to the so called
conformal phase transition (CPT) introduced in Ref. \cite{MY}. There
are the following characteristic features of the CPT:

a) Unlike the conventional Ginzburg-Landau (GL) phase transition, a
parameter governing the phase transition in the CPT is connected with a
marginal (i.e. renormalizable)
operator (in the GL phase transition, such a parameter
is connected with a relevant (i.e. superrenormalizable) operator;
it is usually a mass term).   

b) Though the CPT is a continuous phase transition, there is an
abrupt change of the spectrum of light excitations at a critical point
(line). This is unlike the GL phase transition where the spectrum is
continuous at a critical point (line).

In the present model, the parameter governing the phase transition is
the coupling constant $e$. It is connected with the marginal operator
$j_{\mu}A^{\mu}$. The spectrum of the light excitations is 
discontinuous at the critical line (\ref{crline}). Indeed, in the
subcritical region, with massless fermions, there is a 
Coulomb, conformally invariant, phase describing interactions of
massless fermions and gauge bosons. In the supercritical region,
with massive fermions, there are a lot of bound states, including
$2N_{f}^2$ Nambu-Goldstone bosons corresponding to spontaneous
breakdown of the $U(2N_{f})$ to $U(N_{f}) \times U(N_{f})$. Therefore
these two criterions of the CPT are indeed realized in this model.      

Notice that though $\mbox{QED}_{3+1}$ is a renormalizable theory,
there are new, nonperturbative, divergences in the supercritical
phase (see Eq. (\ref{mass})). These divergences are connected
{\em not} with introducing a 2-brane of vanishing thickness
in the model but with the
strong coupling dynamics. As is well known, such
divergences occur in the strong coupling phase of
$\mbox{QED}_{3+1}$ in the absence of any brane \cite{F,M,B}.
They lead to breakdown of the conformal symmetry (nonperturbative
scale anomaly).

It is instructive to compare the reduced $\mbox{QED}_{3+1}$ with
a 2-brane with the conventional $\mbox{QED}_{2+1}$. The SD
equations in these two models are similar. The difference is
in the form of the gauge field propagator. Instead expression
(\ref{invcharge}), one has \cite{ABKW,ANW}:
\begin{equation}
D(k^2) = \frac{1}{\frac{k^2}{e^2_{3}} + \Pi(k^2)},
\label{invcharge1}
\end{equation}
where $e_{3}$ is the (dimensional) coupling constant in
$\mbox{QED}_{2+1}$
and $\Pi(k^2)$ is the (same) polarization function (\ref{polarization}).
The appearance of the term ${k^2}/{e^2_{3}}$,
instead ${2k}/{e^2}$, makes quite 
a difference. On the one hand, it provides a dynamical
ultraviolet cutoff $\sim e^2_{3}$ in the SD equation and, on the
other hand, since this term is suppressed in the region
$k \ll  e^2_{3}$, it does not contribute to the fermion dynamical
mass. This implies  reducing screening of
Coulomb like interactions as compared to the reduced 
$\mbox{QED}_{3+1}$ with a 2-brane. Indeed,
the dynamical mass in $\mbox{QED}_{2+1}$ is \cite{ANW,nash,QED3}:
\begin{equation}
m_{3d} \sim
e^2_{3}
\exp{\left(-\frac{2\pi}{\nu_{3}}\right)},
\label{3mass}
\end{equation}
where $\nu_{3}=\sqrt{8\lambda_{3}-1}$ with $\lambda_{3}=
16/3{\pi}^{2}N_{f}$. The parameter $\nu_{3}$
coincides with $\nu$ in Eq. (\ref{mass})
only in the limit $e^2 \to \infty$, i.e. in the limit of
maximally strong interactions in the reduced
$\mbox{QED}_{3+1}$ with a 2-brane.\footnote{The critical value
of $N_f$ is $N_{f}^{cr} = 128/3{\pi}^{2} \simeq 4.32$ in
$\mbox{QED}_{2+1}$. Since this result was obtained in the
framework of the $1/N_{f}$ expansion, there may be some concern about its 
reliability \cite{Penn}. Although it would be too strong to
say that this issue has been finally resolved,
different studies
indicate
that $1/N_{f}$ corrections 
are small for $N_f$ around $4$ \cite{nash,QED3}.}

Therefore we conclude that there are important similarities
and important diffirences between the dynamics in
$\mbox{QED}_{2+1}$ and reduced $\mbox{QED}_{3+1}$ with a
2-brane. Both dynamics are intimately connected with 
long range Coulomb like interactions. Both
dynamics provide a realization of the conformal phase
transition. In particular, like in the reduced
$\mbox{QED}_{3+1}$, there is an abrupt change of the
spectrum of light excitations at the critical point
$N_{f} = N_{f}^{cr}$ in $\mbox{QED}_{2+1}$
\cite{ATW}. On the other hand, since
$\mbox{QED}_{2+1}$ is superrenormalizable (and therefore
asymptotically free) theory, there is no (nonperturbative)
ultraviolet divergence in the dynamical mass. Also, its
dynamics is more effective in generating a fermion mass
in that it corresponds to the
dynamics in the reduced $\mbox{QED}_{3+1}$ when the
coupling constant $e$ of the latter goes to $\infty$.

This point is intimately connected with the violation of
the unitarity in the brane theory. Indeed, because of the
first term in the denominator of expression (\ref{invcharge}),
there is an imaginary part for ${\em all}$ time like momenta
$k$ 
in the gauge field propagator
(\ref{full_ph_prop}), ${\em independently}$
of the value of the mass $m_d$. This feature reflects 
the process of escaping of a gauge boson from the brane to
the bulk. This "instability" of brane gauge bosons leads
to an effective reduction of interactions on the brane.
Only in the limit $e \to \infty$ the gauge bosons are 
localized on the brane, i.e. become "stable".    

\section{Reduced $\mbox{QED}_{3+1}$ with a 1-brane}
\label{41}

In this section we will consider the dynamics in the reduced
$\mbox{QED}_{3+1}$ with a 1-brane, i.e. with $D=4$ and
$d=1$. As it was pointed out in Sec. \ref{features}, there
are (classical) ultraviolet divergences in the theory with
a 1-brane of vanishing thickness in this case. Because of that, one
needs
to introduce a finite thickness for the brane,
which will play a role of a regularization parameter.
 
To get the reduction $3+1\rightarrow 1+1$, we perform integration in
Eq.(\ref{action_source}) with the sources taken as
\begin{eqnarray}
J^{a}(X)&=&0 \,\,\, \mbox{for} \,\,\, a = 2,3,\nonumber\\
J^{a}(X)&=&j^{a}(x_4,x_1)f(x_2)f(x_3) \,\,\, \mbox{for} \,\,\, a = 0,1,
\end{eqnarray}
where the regularization function
\begin{equation}
f(x)=\sqrt{\frac{a}{\pi}}\exp{(-ax^2)},\quad f(x)\to\delta(x),
\quad a\to\infty .
\label{regularization}
\end{equation}
Integrating over $x_2,x_3$ and $y_2,y_3$ in Eq.(\ref{action_source}),
we obtain the reduced $1+1$ dimensional action (\ref{redaction})
with the bare gauge field propagator
\begin{eqnarray}
D^{(0)}_{[41]\mu\nu}(x - y)&=&e^2 \int\frac{d^2k dk_2dk_3}{(2\pi)^4}
\exp\left({ik(x-y)}-\frac{k_2^2+k_3^2}{2a}\right)
\left(\delta_{\mu\nu} -\left(1- \xi 
\right)\frac{k_{\mu}k_{\nu}}{ k^2+k_2^2+k_3^2}\right)\frac{1}{
k^2+k_2^2+k_3^2}\nonumber\\
&=&\int\frac{d^2k}{(2\pi)^2} \exp\left({ik(x - y)}\right)
D_{[41]\mu\nu}^0(k),\quad \mu,\nu=4,1,
\end{eqnarray}
where $k=(k_4,k_1),k^2=k_4^2+k_1^2$. It is:
\begin{eqnarray}
D_{[41]\mu\nu}^{(0)}(k)=\left[\delta_{\mu\nu}-
\left(1-\xi \right)\frac{k_{\mu}k_{\nu}}{ k^2}\cdot
k^2\frac{d}{dk^2}\right]D^{(0)}(k^2),
\label{4-1_ph_propag}
\end{eqnarray}
where
\begin{eqnarray}
D^{(0)}(k^2)=-\frac{e^2}{4\pi}\exp{(\frac{k^2}
{2a})}Ei\left(-\frac{k^2}{2a}\right)
\label{D^{(0)}-function}
\end{eqnarray}
and $Ei(-x)$ is the integral exponential function.

By introducing a $1+1$ gauge field $A_{\mu}$, we obtain an effective
$1+1$ dimensional action:
\begin{eqnarray}
S_{[41]eff} &=& \int d^2x \frac{1}{4e^2}F_{\mu\nu}\frac{1}
{-\partial^2 D^{(0)}(-\partial^2)}F^{\mu\nu} + A_{\mu}j^{\mu}
+ \nonumber\\ 
&&
\frac{1}{2e^2}\partial_{\mu}A^{\mu}
\frac{1}{-\partial^2[D^{(0)}(-\partial^2) +
(1 - \xi) \partial^2 D^{(0)^{\prime}}(-\partial^2)]}
\partial_{\nu}A^{\nu},
\label{41effaction}
\end{eqnarray}
where $D^{(0)^{\prime}}
(-\partial^2) = D^{(0)^{\prime}}(x)|_{x=-\partial^2}$.
If there were no need for a regularization, this effective
action would be conformal invariant. The finite
thickness of the 1-brane breaks the conformal symmetry. We will    
return to this point below.

One should add the kinetic term of fermions on a 1-brane
to the action (\ref{41effaction}):
\begin{equation}
S_{kin}= \int d^2x\bar{\psi}\left(i\gamma^{\mu}\partial_{\mu}
\right)\psi.
\label{2fermions}
\end{equation}
We will consider $N_{f}$ two component (i.e. vector like) fermions.
The chiral group is $U(N_{f})_{L} \times U(N_{f})_{R}$.
Our aim is to find whether a fermion dynamical mass is generated
in the theory with effective action (\ref{41effaction}). Naively,
one might expect that in this case the chiral symmetry
$U(N_{f})_{L} \times U(N_{f})_{R}$ would be spontaneously broken down to   
its vector subgroup $U(N_{f})_{V}$. However, this is not the
case in 1+1 dimensions. Due to the
Mermin-Wagner-Coleman (MWC) theorem \cite{MWC}, there cannot be
spontaneous
breakdown of continuous symmetries in $1+1$ dimensions.  The MWC
theorem
is based on the fact that gapless Nambu-Goldstone bosons cannot exist
in $1+1$ dimensions. It however does not prevent a generation
of a fermion mass. In this case, the so called
Berezinski-Kosterlitz-Thouless phase
would realize \cite {BKT,W}. We will
return to this point at the end of this section.

As in the previous section,
we will change the initial gauge in such a way that the
full gauge propagator takes the form:
\begin{eqnarray}
D_{[41]\mu\nu}(k)=\left[\delta_{\mu\nu}-
\left(1-\xi(k^2)\right)\frac{k_{\mu}k_{\nu}}{k^2}\right]D(k^2)
\equiv\left[\delta_{\mu\nu}-
\left(1-\xi(k^2)\right)\frac{k_{\mu}k_{\nu}}{k^2}\right]\frac{1}{D_0^{-1}
(k^2)+\Pi(k^2)}.
\label{4-2_ph_propag_nonlocal}
\end{eqnarray}
Our aim is to find such a function $\xi(k^2)$ that the fermion
wave function $A(k^2)$ would be identically equal to $1$
in the improved ladder approximation. Fortunately, as
Eq. (\ref{gauge}) shows, for $d=1$
the choice of $\xi=1$, i.e. the Feynman
gauge, provides $A(k^2)=1$ for ${\em any}$ gauge field
propagator.
Then the SD equation for the
fermion mass function takes the following form in the improved
ladder approximation:
\begin{eqnarray}
B(p^2)=\frac{2}{(2\pi)^2}\int\frac{d^2kB(k^2)}{k^2+B^2(k^2)}D((p-k)^2).
\label{SDeqmass}
\end{eqnarray}
As was shown in the previous section,
in the reduced $\mbox{QED}_{3+1}$ with a 2-brane, there is no
generation of the fermion mass for a small coupling constant
$\alpha$, and we used the
$1/N_{f}$ expansion in order to justify the use of the
improved ladder approximation. As we will see below, the situation
in the  reduced $\mbox{QED}_{3+1}$ with a 1-brane is different:
there may exist a solution with a nonzero $m_d$ even for an
arbitrary small $\alpha$. For such a $\alpha$, one can expect
that even the ladder approximation is justifiable.

Using the same arguments as in the reduced  $\mbox{QED}_{3+1}$ 
with a 2-brane, one can show that
in the present case the momentum region
yielding a dominant contribution to SD equation (\ref{SDeqmass})
is $m_d \ll k \ll \Lambda$ (we will see below that the
parameter $a^{1/2}$ (the inverse thickness of the 1-brane) plays
the role of an ultraviolet cutoff $\Lambda$). 
Therefore, like in the previous section,     
for the vacuum polarization function 
one can use the one-loop expression with the propagators for free
massive fermions with the mass $m_d$. 
In $1+1$ dimensions, this expression is:
\begin{eqnarray}
\Pi(k^2)=\frac{N_f}{\pi}\left[1-\frac{2m_d^2}{\sqrt{k^2(k^2+4m_d^2)}}
\ln\frac{\sqrt{k^2+4m_d^2}+\sqrt{k^2}}{\sqrt{k^2+4m_d^2}-\sqrt{k^2}}\right].
\label{vacuum_pol}
\end{eqnarray}
The asymptotics of this expression are:
\begin{eqnarray}
&\Pi(k^2)&\longrightarrow\frac{N_fk^2}{6\pi m_d^2},
\quad \mbox{for}\quad
k^2\ll m_d^2,
\label{vacuum_pol1}\\
&\Pi(k^2)&\longrightarrow\frac{N_f}{\pi},\quad\mbox{for}\quad
k^2\gg m_d^2.
\label{vacuum_pol2}
\end{eqnarray}
From Eqs. (\ref{D^{(0)}-function}) and (\ref{vacuum_pol2}) we
find that for momenta $k^2\gg 2a$ the photon propagator
(\ref{4-2_ph_propag_nonlocal}) rapidly decreases:
\begin{equation}
D(k^2)\sim \frac{1}{\frac{4\pi k^2}{2e^2a}+\frac{N_f}{\pi}}\sim   
\frac{2ae^2}{4\pi k^2}.
\end{equation}  
Therefore the whole integrand in SD equation (\ref{SDeqmass})  
rapidly decreases for  $k^2\gg 2a$. Because of that, one 
can neglect the region of those large momenta
and put the cutoff $\Lambda_a^2=2a$ or, more precisely,
$\Lambda_a^2=2a\exp{(-\gamma)}$, with $\gamma$ the Euler constant,
in the SD equation.

This also implies that one can keep the
leading order term in expansion of $D^{(0)}$ (\ref{D^{(0)}-function})
in $k^{2}/a$:
\begin{equation}
D_0(k^2)\simeq \frac{e^2}{4\pi}\log\frac{2a\exp{(-\gamma)}}{k^2} 
\equiv \frac{e^2}{4\pi}\log\frac{\Lambda_a^2}{k^2}.
\label{asbare}
\end{equation}
From  here and 
Eqs. (\ref{4-2_ph_propag_nonlocal}), (\ref{vacuum_pol1}), and
(\ref{vacuum_pol2}) we find that
\begin{eqnarray}
&{D}^{-1}(k^2)&\simeq \frac{4\pi}{e^2\log\frac{\Lambda_a^2}{k^2}},\quad
\mbox{at}\quad k^2\alt m_d^2,\nonumber\\
&{D}^{-1}(k^2)&\simeq \frac{4\pi}{e^2\log\frac{\Lambda_a^2}{k^2}}+
\frac{N_f}{\pi},\quad\mbox{at}\quad k^2\agt m_d^2.
\end{eqnarray}

Now we proceed at solving the SD equation. In order to get a hint of
the character of the
solution, we will first consider the
so-called a constant mass approximation,
taking the external momentum being equal to zero and replacing the running
mass function in the integrand by its value 
$m_d=B(0)$. Then we get the equation
\begin{equation}
1=\int\frac{d^2k}{(2\pi)^2}\frac{1}
{k^2+m_d^2}\frac{2}{\frac{4\pi}{e^2\log
\frac{\Lambda_a^2}{k^2}}+\frac{N_f}{\pi}}.
\label{gap:eq}
\end{equation} 
The main contribution comes from the range of momenta $k^2\gg
m^{2}_{d}$. Therefore
one can omit
the term  $m^{2}_{d}$ in denominator and put instead
the parameter $m^{2}_{d}$
as the lower limit in the integral.
Then the integral can be easily evaluated and one gets the
following algebraic transcendental equation:
\begin{eqnarray}
1=\frac{1}{2N_f}\left\{\log\frac{\Lambda_a^2}{m_d^2}
-\frac{4\pi^2}{e^2N_f}\log\left[
1+\frac{e^2N_f}{4\pi^2}\log\frac{\Lambda_a^2}{m_d^2}\right]\right\}.
\label{gapeq:const_mass}
\end{eqnarray}
Introducing the variable 
$y=(e^2N_f/4\pi^2)\log(\Lambda_a^2/m_d^2)$, it can be
rewritten as
\begin{equation}
\frac{e^2N_f^2}{2\pi^2}=y-\log(1+y).
\label{trans_eq}
\end{equation}
The function on the right hand side of this equation is
monotonically increasing, 
starting from zero value at $y=0$ and 
going to $\infty$ as $y \to \infty$. Therefore this equation
always has a solution. However, the character of the solution depends
on the value of the parameter $e^2N_f^2/2\pi^2$. Indeed, one gets:
\begin{eqnarray}
m_d^2& \simeq &\Lambda_a^2\exp\left(-\frac{4\pi}{e}\right),
\quad \mbox{for}\quad
\frac{e^2N_f^2}{2\pi^2}\ll1,
\label{sol_N_small}\\
m_d^2& \simeq 
&\Lambda_a^2\left(\frac{e^2N_f^2}{2\pi^2}\right)^{-\frac{4\pi^2}
{e^2N_f}}
\exp\left(-2N_f\right),\quad \mbox{for}\quad
\frac{e^2N_f^2}{2\pi^2}\gg1.
\label{sol_N_large}
\end{eqnarray}
Notice that solution (\ref{sol_N_small}) corresponds to
a weak coupling regime. It
does not depend on $N_f$ 
and therefore comes from the range of momenta in the
integral equation where one can neglect the vacuum polarization,
i.e. one can use the ladder (rainbow) approximation in this case.

A closer look at Eq.(\ref{gapeq:const_mass}) reveals that the
solution (\ref{sol_N_small}) emerges from the range of momenta where
a double logarithmic contribution dominates in the gap equation 
(\ref{gap:eq}). On the other hand, the solution at large $N_f$ (Eq. 
(\ref{sol_N_large})) comes from the region of momenta
generating a one logarithmic contribution.  

We will turn now at studying SD equation
(\ref{SDeqmass})
for the running mass function. In order to integrate over
the angle variable there, we will use the following conventional
approximation for the vector boson propagator \cite{HM} (for
a review see Ref. \cite{book}):
\begin{equation}
D((p-k)^2) \simeq  D(p^2)\theta(p^2-k^2)+D(k^2)\theta(k^2-p^2).
\label{D_func_appr}
\end{equation}
This approximation is justifiable:
the measure of the
only "dangerous" (for this approximation) region, with
$|p^2|\simeq |k^2|$, is small and the 
dependence of the propagator $D((p-k)^2)$
on the angular variable is rather smooth.
 
Neglecting then the term $B^2(k^2)$ in the denominator and 
instead putting the infrared cutoff 
$m_d^2\equiv B^2(m_d^2)$ in the integral (the bifurcation
approximation discussed in the previous section),
one gets a simple integral equation. It is easy to show that 
it is equivalent to a differential equation
with two (infrared (IR) and ultraviolet (UV)) boundary conditions:
\begin{equation}
B^{\prime\prime}(x)-\frac{D^{\prime\prime}(x)}{D^{\prime}(x)}B^{\prime}(x)-
\frac{D^{\prime}(x)}{2\pi x}B(x)=0,\quad x=p^2,
\label{eq:diff}
\end{equation}
\begin{equation}
B^\prime(x)|_{x=m_d^2}=0,
\quad \left[D(x)B^{\prime}(x)-D^{\prime}(x)B(x)
\right]\Big|_{x=\Lambda_a^2}=0.
\end{equation}
It is convenient to introduce the variable
$z=(e^2N_f/4\pi^2)\log(\Lambda_a^2/x)$
in terms of which Eq.(\ref{eq:diff}) 
becomes:
\begin{equation}
B^{\prime\prime}(z)-\frac{D^{\prime\prime}(z)}{D^{\prime}(z)}B^{\prime}(z)-
\frac{2\pi}{e^2N_f}D^{\prime}(z)B(z)=0,\quad D(z)=\frac{\pi}{N_f}\frac{z}{z+1},
\label{eq:diffz}
\end{equation}
Together with the boundary conditions, it can be rewritten as
\begin{eqnarray}
&&B^{\prime\prime}(z)+\frac{2}{z+1}B^{\prime}(z)
+\frac{2\pi^2}{e^2N_f^2(z+1)^2}
B(z)=0,\\
&&B^{\prime}(z)\Big|_{z=
\frac{e^2N_f}{4\pi^2}\log\frac{\Lambda_a^2}{m_d^2}}=0,\quad
\left[zB^{\prime}(z)-B(z)\right]\Big|_{z=0}=0.
\end{eqnarray}
The solution $B(z)$ satisfying the UV boundary condition and the
normalization $B(x=m_d^2)=m_d$ is given by
\begin{equation}
B(z)=m_d\left(\frac{z_0+1}{z+1}\right)^{1/2}
\frac{\sinh\left[\frac{\omega}
{2}\log(z+1)\right]}{\sinh\left[\frac{\omega}{2}\log(z_0+1)\right]},\quad
\omega=\sqrt{1-\frac{8\pi^2}{e^2N_f^2}},
\end{equation}
where $z_0 \equiv z(x=m_d^2)$.
The IR boundary condition leads to the equation for the dynamical mass:
\begin{equation}
\tanh\left[\frac{\omega}{2}\log(z_0+1)\right]=\omega.
\label{eq:subcrit}
\end{equation}
For real $\omega$ ($e^2N_f^2/8\pi^2 >1$) it
can be easily solved:
\begin{equation}
m_d^2 \simeq
\Lambda_a^2\exp\left(-N_f\Sigma(\omega)\right),\quad \Sigma(\omega)=
\left[\left(\frac{1
+\omega}{1-\omega}\right)^{1/\omega}-1\right]\frac{1-\omega^2}{2}.
\label{large_N_mass}
\end{equation}
For large $e^2N_f$ this solution becomes
\begin{eqnarray}
m_d^2=\Lambda_a^2\left(\frac{e^2N_f^2}{2\pi^2}
\right)^{-\frac{8\pi^2}{e^2N_f}}
\exp\left(-2N_f\right),
\end{eqnarray}
and coincides, up to minor difference in preexponential factor, with 
expression (\ref{sol_N_large}) obtained in the constant mass
approximation.

The line $\alpha\equiv e^2/4\pi =2\pi/N_f^2$ divides the region in plane
($\alpha,N_f$) in two parts with different dependence of a dynamical
mass on $\alpha$ and $N_f$. Indeed, at $\alpha<2\pi/N_f^2$, 
when $\omega=i\nu, \nu=\sqrt{8\pi^2/e^2N_f^2 -1}$, 
Eq.(\ref{eq:subcrit}) gives 
\begin{eqnarray}
m_d^2\simeq \Lambda_a^2\exp\left(-\frac{\pi^2\sqrt{2}}{e}\right)=
\Lambda_a^2\exp\left(-\pi\sqrt{\frac{\pi}{2\alpha}}\right).
\label{mass_small_e}
\end{eqnarray}
The ratio of powers of two exponents in
(\ref{mass_small_e}) and (\ref{sol_N_small}) is $\pi/2\sqrt2\approx1.11$
that shows that the constant mass approximation in this case is also
rather reliable. 
It is peculiar that the expression (\ref{mass_small_e}) for a dynamical
mass coincides with the expression for a dynamical mass generated by a
magnetic field in quenched 
$\mbox{QED}_{3+1}$ (see Eq. (111) in Ref. \cite{nucl1996}). 
In fact, in the ladder (rainbow) approximation, 
used in the weak coupling regime, the present
SD equation essentially
coincides with the SD equation in that paper (see especially
Appendix C there). The origin of this similarity is in the
dimensional reduction $3+1 \to 1+1$ in the dynamics of
spontaneous chiral symmetry breaking in a magnetic field
\cite{nucl1996}.

The existence of the two types solutions, corresponding to the
weak $(\bar{e}^2 \equiv e^{2}N_{f} \ll 1)$ and the strong 
$(\bar{e}^2 \gg 1)$ coupling regimes, is intriguing. While the 
strong coupling solution essentially coincides with that in the 
$1+1$ dimensional Thirring model (see below),
the weak coupling one yields a new type solution,
characteristic for a 1-brane physics in a $3+1$ dimensional
bulk. These two solutions are generated by very different dynamics: 
while in the strong coupling regime the gauge field propagator is 
dominated by the 1-brane vacuum polarization operator, in the weak 
coupling one the propagator is dominated by the bare term coming 
from the bulk. In particular, while the polarization operator is generated
by the conformal invariant interaction $j_{\mu}A^{\mu}$,
the bare term breaks the conformal symmetry as result of a finite thickness
$1/a^{1/2}$ of a 1-brane. We will argue below that this point can be
important in the
connection with the MWC theorem.  

As we already stated above, there cannot be spontaneous breakdown
of a continuous symmetry in $1+1$ dimensions (the MWC theorem)
\cite{MWC}. This happens because strong fluctuations of would
be NG modes lead to vanishing order parameter connected with such
a breakdown. 
Let us recall how this theorem is realized in the
case of the $1+1$ dimensional Thirring model with the color group
$U(N_c)$ and the chiral symmetry $U(N_{f})_{L} \times U(N_{f})_{R}$.
It is relevant for our case since the dynamics of the
strong coupling solution found above essentially
coincides with the dynamics of the Thirring model with the
color group $U(1)$. 
Indeed, since for this solution the gauge field
propagator is dominated by the 1-brane vacuum polarization
function, which is essentially constant in this case (see
Eq. (\ref{vacuum_pol2})), the interaction is of a 
current $\times$ current form, as in the Thirring model.    

First of all, using the Fierz
identities, it is easy to show that, in $1+1$ dimensions, the
Thirring model is equivalent to the Gross-Neveu (GN) model
\cite{GN}. The interaction term of the latter is:
\begin{equation}
S_{int}^{GN}= \int d^2x
\frac{G}{2} 
\left[ (\bar{\psi}\lambda^{s}\psi)^2+
(\bar{\psi}\lambda^{s}i\gamma^5\psi)^2 \right],
\label{gn}
\end{equation}
where ${\lambda}^{s}$ are flavor matrices,
$s=0,1,...,N_f^{2}-1$,
and the summation over $s$ and color indices of the fermion fields is 
assumed.
The ${\lambda}^{s}$ matrices are normalized according to
$\mbox{tr}({\lambda}^{s}{\lambda}^{k})=2\delta^{sk}$.

Let us first consider the case of the 
$U(1)_{L} \times U(1)_{R}$ chiral group. In this case the model is
soluble \cite{AL}. There is a nonzero dynamical mass for fermions for
all $N_c \geq 2$. However, there is ${\em no}$ NG  boson in the
model. Instead of that, there is a Berezinski-Kosterlitz-Thouless (BKT)
gapless mode. This mode is described by the exponent field
$U(x)=\exp(i\theta(x))$, where $\theta$ satisfies the constraint
$0 \leq \theta(x) <2\pi$. More precisely, the BKT mode is 
described by a usual Lagrangian density of a massless free field,
${f}/{2}(\partial_{\mu}\theta \partial^{\mu}\theta)$
with $f\simeq N_{c}/4\pi$.
However, the corresponding 
observables are described not by Green's functions of the
field $\theta$ but by Green's functions of the
field $U(x)$ and its derivatives, including the derivative
$\partial_{\mu}\theta= iU\partial_{\mu}U^{\dag}$. The point is
that while the propagator and other Green's functions
of the $\theta(x)$ field do not exist in $1+1$ dimensions
(they are divergent for all $x$), Green's functions of the $U(x)$
field 
and its derivatives are well defined. Moreover, the corresponding
field theory is conformal invariant and the parameter $f$ 
defines anomalous dimensions of its Green's functions.

The case of the GN model with one color is special. For 
$N_{c}=1$ and the
chiral group $U(1)_{L} \times U(1)_{R}$, fermions are massless
and, moreover, the bosonization of the model leads exactly
to the Lagrangian of the free massless BKT mode \cite{C}. 
Therefore
in this particular case, the ${\em whole}$ dynamics is conformal
invariant.

Though the dynamics with $N_f \geq 2$ is more involved, some of the
basic points described above survive. In this case one should
distinguish the $U(1)_{L} \times U(1)_{R}$ and the
$SU(N_f)_{L} \times SU(N_f)_{R}$ sectors. The dynamics in the 
first one is essentially the same as in the model with $N_{f}=1$
and one should expect
that
while for $N_{c} \geq 2$ fermions are massive, they become
massless for $N_{c}=1$. In the second sector, because of 
a strong self-interaction between $N_{f}^2-1$ would be NG bosons,
all of them 
acquire a (same) mass, thus leading to a Wigner 
realization of the dynamics with the exact
$SU(N_f)_{L} \times SU(N_f)_{R}$ symmetry \cite{PW}.

In our case the number of colors $N_{c}=1$. Does it necessarily
imply that the dynamical mass of fermions will disappear in the
exact solution in the reduced $\mbox{QED}_{3+1}$
with a 1-brane? 
We do not think that the situation is so simple.
First of all, even the status of the Goldstone theorem is not
completely clear in this case: some of the assumptions 
the theorem is
based on are violated in the brane world. Indeed, in
the initial bulk theory, the $(D-1)+1$ Lorentz invariance is
broken because of the presence of a d-brane. On the other hand, while 
on a d-brane the $d+1$ Lorentz symmetry is
preserved, the corresponding effective theory in nonlocal.
Second, as was emphasized above, in the $1+1$ dimensional
Thirring (or Gross-Neveu) model, it is important that
the conformal symmetry is exact in the sector with the
BKT field $U(x)$. On the other hand, in the reduced     
$\mbox{QED}_{3+1}$ with a 1-brane, the conformal symmetry is
necessarily broken by a finite thickness of the brane. The
latter is especially important for the weak coupling solution
(\ref{mass_small_e}) in which the gauge field propagator is
dominated by the the bare term (\ref{asbare}) which 
explicitly breaks the conformal symmetry. The dynamics described by 
that solution is very different from that of the Thirring model.

It remains a challenge to clarify these various issues in the brane
dynamics.

\section{Reduced $\mbox{QED}_{2+1}$ with a 1-brane}
\label{31}

In this section we will study spontaneous chiral symmetry breaking
in the reduced $\mbox{QED}_{2+1}$ with a 1-brane, i.e. with D=3 and d=1.
Recall that the gauge coupling constant is dimensional in $2+1$
dimensions: its dimension is $[e_{3}]=m^{\frac{1}{2}}$, and we will
see that the parameter $e_{3}^{2}N_{f}$ plays the role of an
ultraviolet cutoff, which is a typical feature for $\mbox{QED}_{2+1}$ 
\cite{P,ABKW}. Notice also that as it follows from the discussion in 
Sec. \ref{features}, there is no need for introducing a finite
thickness for a 1-brane in a $2+1$ dimensional bulk.

With trivial modifications, the effective action can be derived as in the
case of the reduced $\mbox{QED}_{3+1}$ with a 2-brane. It is:
\begin{equation}
S_{[32]eff} = \int d^2x \left[ \frac{1}{2e_{3}^2}F_{\mu\nu}
\frac{1}{\sqrt{-\partial^2}}F^{\mu\nu} + 
\bar{\psi}(i\gamma^{\mu}\partial_{\mu})\psi +
A_{\mu}j^{\mu} +\frac{1}{e_{3}^2\xi}\partial_{\mu}A^{\mu}
\frac{1}{\sqrt{-\partial^2}}\partial_{\nu}A^{\nu} \right].
\end{equation}
As in the previous section, we 
will consider $N_f$ two component fermion fields
(see Eq. (\ref{2fermions})).
The chiral group is $U(N_f)_{L} \times U(N_f)_{R}$. 

The full photon
propagator in a nonlocal gauge is given by
\begin{equation}
D_{\mu\nu}(k)=\left[\delta_{\mu\nu}-
\left(1-\xi(k^2)\right)\frac{k_{\mu}k_{\nu}}{k^2}\right]\frac{1}
{\frac{2k}{e_{3}^2}+\Pi(k^2)},
\label{32_ph_propag_nonlocal}
\end{equation}
where the vacuum polarization function $\Pi(k^2)$ 
is given in Eq.(\ref{vacuum_pol}). As was
shown in Sec. \ref{41},
the convenient choice of the gauge for the study of
fermion dynamics on a 1-brane is $\xi=1$. Then the SD equation
for the fermion mass function takes the
form of Eq.(\ref{SDeqmass}) with the function $D(k)$ given now by
\begin{equation}
D(k)=\frac{1}{\frac{2k}{e_{3}^2}+\Pi(k^2)}.
\label{D_32func}
\end{equation}
Like in the case of the reduced $\mbox{QED}_{3+1}$ with a
1-brane, the bare term $2k/e_{3}^{2}$ breaks the conformal
symmetry. However, this bare term is very different from that
one in Eq. (\ref{gap:eq}). Its strong dependence on momentum implies
that it is important both in the infrared and ultraviolet regions.   
Taking into account this term and  
the asymptotics of the vacuum polarization function $\Pi(k^2)$
(\ref{vacuum_pol1}) and (\ref{vacuum_pol2}), one concludes
that the dominant, logarithmic, 
contribution to the SD equation should 
come from the range of momenta
$12\pi m_d^{2}/e_{3}^2N_f<k<e_{3}^{2}N_f/2\pi$.

An analysis of this SD equation is done in Appendix A. It is shown
there that a solution with a nonzero dynamical mass exists for all 
values of $N_f$ and $e_{3}^2$. It is also shown that the 
dynamical mass satisfies the following constraint:
\begin{equation}
\frac{N_{f}e_{3}^{2}}{\pi} \exp(-2N_f) \alt  m_d \alt 
\frac{N_{f}e_{3}^{2}}{2\sqrt{6}\pi}
\exp\left(-\frac{N_f}{7}\right).
\label{constraint}
\end{equation}
In the case when $N_f \gg 1$, 
the dynamical mass is:
\begin{equation}
m_{d}\simeq \frac{N_fe_{3}^2}{2\sqrt{6}\pi}
\exp[-(N_f+\frac{1}{8}+\gamma-3\log7)].
\label{31mass}
\end{equation}
It is interesting that, unlike the previous model with $D=4$ and
$d=1$, in this model the constant mass approximation is unreliable. 
In particular, it is not difficult to show (see Appendix A) that it
would yield the following expression for the dynamical mass in the 
case $N_{f} \gg 1$:
\begin{equation}
m_d\simeq \frac{N_fe_{3}^2}{2\sqrt{6}\pi}
\exp\left(-\frac{N_f}{7}\right),
\label{const_mass_approx}
\end{equation}     
which is very different from expression (\ref{31mass}).
The reason for that is that, unlike the previous model, the bare
term in the propagator (\ref{D_32func}) (now strongly depending on
momentum) does not decouple even in the dynamical regime with $N_{f} \gg
1$.   

We would like also to add that all remarks made in Sec. \ref{41} 
concerning the status of the problem of the fermion mass generation on 
a 1-brane, in particular, its connection with the Mermin-Wagner-Coleman 
theorem, are also relevant for the present case. 

\section{Conclusion}
\label{conclusion}

The dynamics of chiral symmetry breaking in reduced 
$\mbox{QED}$ is rich and quite nontrivial. Its characteristic
features are intimately connected with the structure of the
gauge field propagator. It includes two terms: the vacuum
polarization function, completely defined by the brane
dynamics, and the "bare" term coming form the bulk. The vacuum
polarization function is connected with the conformal
invariant term $j_{\mu}A^{\mu}$. Therefore, since in $1+1$ and
$2+1$ dimensions there are no divergences in
the polarization function, it is conformal invariant for
massless fermions. 
This feature essentially survives
in the near-critical regime of chiral symmetry breaking: in
this regime, a fermion dynamical mass $m_d$ is small and
the dominant region is that with momenta $k \gg m_d$. 

On the other hand, in many cases, the bare term breaks the
conformal symmetry: either because of a finite thickness
of a brane or because an initial bulk theory (as
$\mbox{QED}_{2+1}$) is not conformal invariant. The
interplay between those two dynamical sources provides 
rich nonperturbative dynamics.

In this paper, the improved rainbow approximation (with a bare
vertex) was used. It
would be interesting to study the dynamics beyond this approximation,
though it is not straightforward at all.
The point is that, besides the bare spin structure $\gamma_{\mu}$, the  
vertex can have other ones. For example, in the case of a 2-brane, there
are in principle $11$ other structures. The crucial point in the 
present analysis
is decoupling of the Schwinger-Dyson equations. Therefore the role of   
the gauge where the function $A(p^2)=1$ is very important. As it is
discussed in Sec. \ref{42}, for the bare vertex, one can find
such a gauge for any vector boson propagator. However, beyond the  
approximation with the bare vertex, new structures in the vertex can
appear. In order to find these structures, one 
either should consider the equation for the vertex (that is quite
complicated) or try to construct
an ansatz for the vertex consistent with such general constraints as
Ward identities, the absence of kinematic singularities, the 
correct perturbative limit, etc.. This last
approach was successful in 3+1 dimensional QED \cite{CP}.
However, studies of this problem 
in 2+1 dimensional QED (which is similar to the 
dynamics on a 2-brane) have revealed that it is a hard (and still
unresolved) problem \cite{BKP}. We hope to turn to this problem
elsewhere.

At last, we would like to indicate that this analysis can be useful for
studying dynamical chiral symmetry breaking in higher dimensional
brane theories \cite{D,HTY}. In this connection, it is
noticeable that in Ref. \cite{HTY} the consequences of
the existence of a ultraviolet stable fixed point in
higher dimensional gauge theories were
considered.
\acknowledgements{
E.V.G. would like to thank the members of Physics Department of Nagoya
University, Japan, and the
International Center for Theoretical Physics, Trieste, Italy, for
financial support and hospitality during his stay there.
This work is partially supported by
Grant-in-Aid of Japan Society for the Promotion of Science (JSPS)
\#11695030.
The work of V.P.G. was also supported by the SCOPES grant 7 IP 062607  
of the Swiss NSF. He wishes to
acknowledge JSPS for financial support.}

\appendix

\section{Analysis of the Gap Equation for the Reduced 
$\mbox{QED}_{2+1}$ with a
1-brane}

In this Appendix we analyse the SD equation for the case
of the $\mbox{QED}_{2+1}$ with a 1-brane.
The equation has the form
\begin{equation}
B(p^2)=\frac{2}{(2\pi)^2}\int\frac{d^2kB(k^2)}{k^2+B^2(k^2)}
D((p-k)^2),
\label{A32SDeq:mass}
\end{equation}
where
\begin{eqnarray*}
D((p-k)^2) = 
\frac{1}{\frac{2\sqrt{(p-k)^2}}{e_{3}^2}+\Pi((p-k)^2)},
\end{eqnarray*}
and the vacuum polarization function $\Pi(k^2)$
is given in Eq. (\ref{vacuum_pol}).

We will first obtain the constraint (\ref{constraint}) for the
dynamical mass. We begin by deriving the lower limit for $m_d$.
As was already indicated in Sec. \ref{31}, 
the dominant contribution to SD equation (\ref{A32SDeq:mass}) comes
from the range of momenta $\mu<k<\Lambda$ with infrared and
ultraviolet cutoffs given by
$\mu=12\pi m_d^{2}/N_fe_{3}^2$, $\Lambda=e_{3}^{2}N_f/2\pi$.
Since the kernel of this integral equation
is positive (corresponding to an attractive interaction),
we obviously obtain a lower limit 
for $m_d$ if integrate only over this range of momenta
and, furthermore,
replace $D((p-k)^2)$ in
the kernel by its minimal value in this
interval. Taking into account Eq. (\ref{D_func_appr}), one finds
that the minimal value is $\frac{\pi}{2N_f}$.
Then the gap equation becomes simple:
\begin{equation}
B(p^2)=\frac{2}{(2\pi)^2}\int_\mu^\Lambda\frac{d^2kB(k^2)}{k^2+B^2(k^2)}
\frac{\pi}{2N_f}.
\label{ANJL-like}
\end{equation}  
It has the following solution:
\begin{equation}
B(p^2)= m_d \simeq \frac{e_{3}^{2}N_f}{\pi} \exp(-2N_f).
\label{frombelow}
\end{equation} 
Since the initial
interaction is stronger, the
true $m_d$ is larger than the value (\ref{frombelow}).

Let us find an estimate from above for the dynamical mass. To do this, we
consider the integral equation at $p^2=0$. It is
\begin{equation}
B(0)=\frac{2}{(2\pi)^2}\int\frac{d^2kB(k^2)}{k^2+m_d^2}
D(k^2,m_d^2),
\label{Aatzero}
\end{equation}
and we explicitly indicated the dependence of the interaction kernel
$D(k^2,m_d^2)$ on the dynamical mass $m_d^2$.
Eq.(\ref{Aatzero}) is equivalent to
\begin{equation}
1=\frac{2}{(2\pi)^2}\int\frac{d^2kf(k^2)}{k^2+m_d^2}
D(k^2,m_d^2),
\label{Af}
\end{equation}
where $f(k^2) = \frac{B(k^2)}{B(0)}$. It follows from the gap equation that
$B^{\prime}(p^2) < 0$, i.e., $B(p^2)$ is a decreasing function of $p^2$.
Therefore, $f(k^2) < 1$ in Eq.(\ref{Af}) for $k^2 > 0$.
In the case of the constant mass approximation
(where $B(k^2)$ is a constant $B(k^2)=M$) the square mass $M^2$ satisfies
the following gap equation:
\begin{equation}
1=\frac{2}{(2\pi)^2}\int\frac{d^2k}{k^2+M^2}
D(k^2,M^2).
\label{Aconstantmassapprox}
\end{equation}
By using asymptotics (\ref{vacuum_pol1}) and (\ref{vacuum_pol2})
for the vacuum polarization function, the integration region in 
Eq.(\ref{Aconstantmassapprox}) is divided into two
regions $k \alt M\sqrt{6}$ and $k \agt M\sqrt{6}$
\footnote{The value
${M\sqrt{6}}$ here was
determined from matching small and large $k$ asymptotics of the
vacuum polarization function.}
that gives us the following gap equation:
\begin{eqnarray}
1=\frac{1}{\pi}\left[\int\limits_0^{M\sqrt{6}}
\frac{dk}{k^2+M^2}\frac{1}{\frac{2}{e_{3}^2}
+\frac{N_f}{6\pi}\frac{k}{M^2}}+\int\limits_{M\sqrt{6}}^\infty
\frac{dk}{k^2+M^2}\frac{k}
{\frac{2k}{e_{3}^2}+\frac{N_f}{\pi}}\right].
\label{Agapeq:32}
\end{eqnarray}
We can further neglect $k^2$ term in
comparison to $M^2$ in the first integral in Eq.(\ref{Agapeq:32}),
while
in the second one we can neglect $M^2$ in comparison to $k^2$.
Evaluating the integrals, we come to the following expression:
\begin{equation}
1=\frac{6}{N_f}\log\left(1+\frac{N_fe_{3}^2}
{2\sqrt{6}\pi M}\right)+\frac{1}{N_f}\log\left(1+
\frac{N_fe_{3}^2}{2\sqrt{6}\pi M}\right).
\end{equation}
The corresponding solution for a small
dynamical mass ($M\ll e_{3}^2$) is:
\begin{equation}
M\simeq \frac{N_fe_{3}^2}{2\sqrt{6}\pi}\exp\left(-\frac{N_f}{7}\right).
\label{Aconst_mass_approx}
\end{equation}
It is obviously valid for $N_f\gg1$.

Let us prove that $m_d^2 < M^2$, where $m_d^2$ is the solution
of the gap equation with the running
mass function, by assuming the opposite and then showing that it
leads to a contradiction.

So let us assume that $m_d^2 > M^2$ and consider the
integral
\begin{equation}
\int\frac{d^2kf(k^2)}{k^2+m_d^2}D(k^2,m_d^2).
\end{equation}
Since $f(k^2) < 1$, we have
\begin{equation}
\int\frac{d^2kf(k^2)}{k^2+m_d^2}D(k^2,m_d^2) < 
\int\frac{d^2k}{k^2+m_d^2}D(k^2,m_d^2).
\end{equation}
By calculating
\begin{eqnarray*}
I(m_{d}^2) = \int\frac{d^2k}{k^2+m_d^2}D(k^2,m_d^2),
\end{eqnarray*}
one can show
that $I^{\prime}(m_d^2) < 0$, i.e., $I(m_d^2)$ is a decreasing function of
$m_d^2$. Since we assumed that $m_d^2 > M^2$, we have
\begin{eqnarray*}
\int\frac{d^2k}{k^2+m_d^2}D(k^2,m_d^2) < 1
\end{eqnarray*}
and, consequently, we obtain that
\begin{equation}
\int\frac{d^2kf(k^2)}{k^2+m_d^2}D(k^2,m_d^2) < 1
\end{equation}
for all $m_d^2 > M^2$. Then, 
since we cannot satisfy equation (\ref{Af}) with $m_d^2
> M^2$,
the assumption that
$m_d^2 > M^2$ leads to a contradiction. Therefore, 
we get the inequality $m_d^2 < M^2$ with $M$ given in
Eq. (\ref{Aconst_mass_approx}).
This and the lower limit we 
obtained earlier lead us to constraint (\ref{constraint}).
 
Can one get an explicit solution of the integral equation
(\ref{A32SDeq:mass}) in a reliable approximation? The answer is
affirmative.

To solve Eq.(\ref{A32SDeq:mass}),
we use approximation (\ref{D_func_appr}) in order to be able to
perform integration over angles. Since we already know that
the main
(logarithmic) contribution comes from the range of momenta $12\pi 
m_d^2/N_fe_3^2<k<e_3^2N_f/2\pi$, we put infrared and ultraviolet
cutoffs in the integral equation at $\mu=12\pi m_d^2/N_fe_3^2$ and 
$\Lambda=e_3^2N_f/2\pi$, respectively. Then the integral 
equation (\ref{A32SDeq:mass}) takes the form:
\begin{equation}
B(p)=\frac{1}{\pi}\left[D(p)\int\limits_{\mu}^p\frac{dkkB(k)}{k^2+m_d^2}+
\int\limits_p^{\Lambda}\frac{dkkB(k)D(k)}{k^2+m_d^2}\right].
\label{Adiffeq_32}
\end{equation}
Furthermore, we approximate the function $D(p)$ on the interval
$\mu<p<\Lambda$ as 
\begin{equation}
D(p)=\theta(p_m-p) \frac{6\pi m_d^2}{N_fp^2} +
\theta(p-p_m) \frac{1}{\frac{2p}{e_3^2}+\frac{N_f}{\pi}},
\end{equation}
where the parameter $p_m$ is determined from the condition of continuity
of $D(p)$ at the point $p=p_m$, ($p_m=\frac{6\pi m_d^2}{N_fe_3^2} +
\sqrt{(\frac{6\pi m_d^2}{N_fe_3^2})^2 + 6m_d^2}\approx m_d\sqrt{6}$).

It is convenient to represent the mass function as 
\begin{equation}
B(p)\equiv B_i(p)\theta(p_m-p)+B_u(p)\theta(p-p_m).
\end{equation}
For ``infrared'' $B_i$ and ``ultraviolet'' $B_u$ (with respect to
the parameter $p_m$) parts of the mass function, we get the 
following equations:
\begin{eqnarray}
B_i(p) = \frac{6m_d^2}{N_fp^2}\int\limits_{\mu}^{p}
\frac{dkkB_i(k)}{k^2+m_d^2}+\int\limits_{p}^{p_m}
\frac{dkkB_i(k)}{k^2+m_d^2}\cdot\frac{6m_d^2}{N_fk^2} + 
\frac{1}{\pi}\int\limits_{p_m}^{\Lambda}\frac{dkkB_u(k)}{k^2+m_d^2}
\cdot\frac{1}{\frac{2k}{e_3^2}+\frac{N_f}{\pi}},\label{AB_i:inteq}
\end{eqnarray}
\begin{eqnarray}
B_u(p) = \frac{1}{\pi}\left[\frac{1}{\frac{2p}{e_3^2}+\frac{N_f}{\pi}}\int\limits_{p_m}^p
\frac{dkkB_u(k)}{k^2+m_d^2}+
\int\limits_p^{\Lambda}
\frac{dkkB_u(k)}{k^2+m_d^2}\frac{1}{\frac{2k}{e_3^2}+\frac{N_f}{\pi}}\right] +
\frac{1}{\pi}\frac{1}{\frac{2p}{e_3^2}+\frac{N_f}{\pi}}\int\limits_{\mu}^{p_m}
\frac{dkkB_i(k)}{k^2+m_d^2}.
\label{AB_u:inteq}
\end{eqnarray}
Taking the derivatives on the both sides of these equations, we get:
\begin{eqnarray}
B_i^\prime(p)&=&\frac{1}{\pi}\left[-\frac{12\pi m_d^2}{N_fp^3}\int\limits_{\mu}^{p}
\frac{dkkB_i(k)}{k^2+m_d^2}\right],\label{AB_i_prime}
\\
B_u^\prime(p)&=&-\frac{1}{\pi}\frac{2}{e_3^2}\frac{1}{(\frac{2p}{e_3^2}+\frac{N_f}{\pi})^2}
\left[\int\limits_{\mu}^{p_m}\frac{dkkB_i(k)}{k^2+m_d^2}+\int\limits_{p_m}^p
\frac{dkkB_u(k)}{k^2+m_d^2}\right].
\label{AB_u_prime}
\end{eqnarray}
Differentiating the last equations once more time we obtain
\begin{eqnarray}
B_i^{\prime\prime}(p)+\frac{3}{p}B_i^{\prime}(p)+\frac{12m^2}{N_f}\frac{B(p)}{p^2(p^2
+m_d^2)}=0,
\label{Adiffeq_small_mom}\\
B_u^{\prime\prime}(p)+\frac{2}{p+\frac{e_3^2N_f}{2\pi}}B_u^\prime(p)+\frac{e_3^2}{2\pi}
\frac{pB_u(p)}{(p+\frac{e^2N_f}{2\pi})^2(p^2+m_d^2)}=0.
\label{Adiffeq_large_mom}
\end{eqnarray}
We have also the following IR and UV boundary conditions:
\begin{equation}
B_i^\prime(p)\Big|_{p=\mu}=0,\quad\left[
\left(p+\Lambda\right){B_u(p)}\right]^\prime\Big|_{p=\Lambda}=0,
\label{Aboundary_cond}
\end{equation}
where the prime denotes the derivative with respect to $p$. Furthermore,
the mass function is continuous at the point $p_m$, therefore,
$B_i(p_m)=B_u(p_m)$ and the first derivatives satisfy
\begin{equation}
B_u^\prime(p_m)=\frac{6\pi m_d^2}{N_fe_3^2p_m}B_i^\prime(p_m)
\end{equation}
(the condition of continuity
and (A23) follow from Eqs.(\ref{AB_i:inteq}),
(\ref{AB_u:inteq}) and Eqs.(\ref{AB_i_prime}), (\ref{AB_u_prime}),
respectively).

The general solution of Eq.(\ref{Adiffeq_small_mom}) is given in terms
of hypergeometric functions:
\begin{eqnarray}
B(p^2)&=&{C_1}\left(\frac{m_d^2}{p^2}\right)^{\frac{1-\omega}{2}}
F\left(\frac{-1 + \omega}{2},\frac{1 +\omega}{2},1+\omega; -\frac{p^2}
{m_d^2}\right)\nonumber\\
&+&{C_2}\left(\frac{m_d^2}{p^2}\right)^{\frac{1+\omega}{2}}
F\left(-\frac{1 + \omega}{2},\frac{1 -\omega}{2},1-\omega; -\frac{p^2}
{m_d^2}\right),
\end{eqnarray}
where $\omega=\sqrt{1-\frac{12}{N_f}}$.
The IR boundary condition gives a relation between the constants $C_1$
and $C_2$
\begin{eqnarray}
C_1(1-\omega)&\left(\frac{\mu}{m_d}\right)^\omega& F\left[\frac{1+\omega}{2},
\frac{1+\omega}{2};1+\omega;-\left(\frac{\mu}{m_d}\right)^2\right]+
C_2(1+\omega)\nonumber\\
\times&\left(\frac{\mu}{m_d}\right)^{-\omega}& F\left[\frac{1-\omega}{2},
\frac{1-\omega}{2};1-\omega;-\left(\frac{\mu}{m_d}\right)^2\right]=0,
\label{AIR_bound_cond}
\end{eqnarray}
where we used the formula for differentiating the hypergeometric
function \cite{BE} 
\begin{equation}
\frac{d^n}{dz^n}\left[z^{a+n-1}F\left(a,b;c;z\right)\right]=(a)_nz^{a-1}
F\left(a+n,b;c;z\right).
\end{equation}
Since for $B_u(p)$ the corresponding momenta are
larger than $m_d$ ($p\ge p_m>m_d$),
we approximate $p^2 + m_d^2$ by $p^2$ in
Eq.(\ref{Adiffeq_large_mom}). This gives us:
\begin{equation}
B_u^{\prime\prime}(p)+\frac{2}{p+\frac{e_3^2N_f}{2\pi}}B_u^\prime(p)+\frac{e_3^2}{2\pi}
\frac{B_u(p)}{p(p+\frac{e_3^2N_f}{2\pi})^2}=0.
\label{AB_u-diffeq}
\end{equation}
Introducing the variable $z=-2\pi p/N_fe_3^2$ and making the
substitution 
\begin{equation}
B_u(z)=f(t),  \quad \frac{z}{z-1}=t,
\end{equation}
Eq. (\ref{AB_u-diffeq}) reduces to the hypergeometric
differential equation
\begin{equation}
t(1-t)f^{\prime\prime}(t)+\frac{1}{N_f}f(t)=0.
\end{equation}
A solution regular at zero is
\begin{equation}
f_1(t)=tF\left(\frac{1+\nu}{2},\frac{1-\nu}{2};2;t\right),\quad
\nu=\sqrt{1+\frac{4}{N_f}},
\end{equation}
and the second independent solution is \cite{BE}
\begin{equation}
f_2(t)=(1-t)F\left(\frac{1+\nu}{2},\frac{1-\nu}{2};
2;1-t\right).
\end{equation}
Thus, the general solution for the mass function is
\begin{equation}
B_u(z)=C_3\frac{z}{1-z}F\left(\frac{1+\nu}{2},\frac{1-\nu}{2};
2;\frac{z}{z-1}\right)+C_4\frac{1}{1-z}F\left(\frac{1+\nu}{2},\frac{1-\nu}{2};
2;\frac{1}{1-z}\right).
\label{Asol_two}
\end{equation}
The UV boundary condition (\ref{Aboundary_cond}), which can be
rewritten as
\begin{equation}
\frac{d}{dz}\left[(1-z)B(z)\right]\Big|_{z=-1}=0,
\end{equation}
allows us to fix the ratio $C_3/C_4$
\begin{equation}
\frac{C_3}{C_4}=\frac{F\left(\frac{1+\nu}{2},\frac{1-\nu}{2};
2;\frac{1}{2}\right)-F\left(\frac{1+\nu}{2},\frac{1-\nu}{2};
1;\frac{1}{2}\right)}{F\left(\frac{1+\nu}{2},\frac{1-\nu}{2};
1;-\frac{1}{2}\right)+F\left(\frac{1+\nu}{2},\frac{1-\nu}{2};
2;-\frac{1}{2}\right)},
\label{AratioC3_C4}
\end{equation}
where the formula for differentiating the hypergeometric function
\begin{equation}
\frac{d^n}{dz^n}\left[z^{c-1}F\left(a,b;c;z\right)\right]=(c-n)_nz^{c-1-n}
F\left(a,b;c-n;z\right)
\label{Adiff_hyper}
\end{equation}
has been used. For $N_f \gg 1$ we have $C_3/C_4\simeq 1/8N_f$.

Finally, matching the solutions $B_i(p)$ and $B_u(p)$ at the point
$p=p_m\simeq m\sqrt{6}$, we obtain two
other equations for the constants
$C_1,C_2,C_3,C_4$:
\begin{eqnarray}
&&C_16^{\frac{\omega-1}{2}}F\left(\frac{-1+\omega}{2},\frac{1+\omega}{2};1+\omega;-6\right)+
C_26^{-\frac{1+\omega}{2}}F\left(-\frac{1+\omega}{2},\frac{1-\omega}{2};1-\omega;-6\right)=
\nonumber\\
&&\left[C_3\frac{z}{1-z}F\left(\frac{1+\nu}{2},\frac{1-\nu}{2};
2;\frac{z}{z-1}\right)+C_4\frac{1}{1-z}F\left(\frac{1+\nu}{2},\frac{1-\nu}{2};
2;\frac{1}
{1-z}\right)\right]\Big|_{z=-\frac{\mu}{p_m}} \,\,\,\,\,\,\,\,\,\, ,
\label{Amatching1}
\end{eqnarray}
\begin{eqnarray}
&&C_1(1-\omega)6^{\frac{\omega-1}{2}}F\left(\frac{1+\omega}{2},\frac{1+\omega}
{2};1+\omega;-6\right)+C_2(1+\omega)6^{-\frac{1+\omega}{2}}
F\left(\frac{1-\omega}{2},\frac{1-\omega}{2};1-\omega;-6\right)\nonumber\\
&&=\frac{1}{3(1-z)^2}\left[C_3F\left(\frac{1+\nu}{2},\frac{1-\nu}{2};
1;\frac{z}{z-1}\right)+C_4F\left(\frac{1+\nu}{2},\frac{1-\nu}{2};
1;\frac{1}{1-z}\right)\right]\Big|_{z=-\frac{\mu}{p_m}}.
\label{Amatching2}
\end{eqnarray}
The determinant of the set of homogeneous equations
(\ref{AIR_bound_cond}),(\ref{AratioC3_C4}),
(\ref{Amatching1}), and (\ref{Amatching2}) gives an equation for determining
the dynamical mass. Since we
look for a solution with $\frac{\mu}{p_m} \ll1$, we can
simplify the equations for $C_i$ by using the corresponding
formulas for hypergeometrical functions \cite{BE}. Finally, we
obtain the following equation for the dynamical mass:
\begin{eqnarray}
\sqrt{1-\omega^2}A(\omega)\frac{\sinh\left[\omega\left(\log\frac{p_m}{\mu}+
\delta_1(\omega)\right)\right]}
{\sinh\left[\omega\left(\log\frac{p_m}{\mu}+\delta_2(\omega)\right)\right]}=
\frac{1}{3}\frac{C_3}{C_4}\Gamma(\frac{3-\nu}{2})
\Gamma(\frac{3+\nu}{2})-\frac{1}{3N_f}\log\left(\frac{p_m
\exp{(h_0^{\prime\prime})}}
{\mu}\right),
\label{Agapeq_small_m}
\end{eqnarray}
where 
\begin{equation}
A(\omega)=\left[\frac{F\left(\frac{1+\omega}{2},\frac{1+\omega}{2};1+\omega;
-6\right)F\left(\frac{1-\omega}{2},\frac{1-\omega}{2};1-\omega;
-6\right)}{F\left(\frac{-1+\omega}{2},\frac{1+\omega}{2};1+\omega;
-6\right)F\left(-\frac{1+\omega}{2},\frac{1-\omega}{2};1-\omega;-6\right)}\right]^{1/2},
\end{equation}
\begin{eqnarray}
  \delta_1(\omega)&=&\frac{1}{2\omega}\log\frac{F\left(\frac{1+\omega}{2},
\frac{1+\omega}{2};1+\omega;
-6\right)}{F\left(\frac{1-\omega}{2},\frac{1-\omega}{2};1-\omega;-6\right)},\\
\delta_2(\omega)&=&\frac{1}{2\omega}\log\frac{(1+\omega)F\left(\frac{-1+\omega}{2},
\frac{1+\omega}{2};
1+\omega;-6\right)}{(1-\omega)F\left(-\frac{1+\omega}{2},\frac{1-\omega}{2};
1-\omega;-6\right)},
\end{eqnarray}
and the constant $h_0^{\prime\prime}$ is
\begin{eqnarray}
h_0^{\prime\prime}=2\psi(1)-\psi\left(\frac{1+\nu}{2}\right)-\psi\left(\frac{1-\nu}{2}\right).
\end{eqnarray}
In the limit of large $N_f \gg 1$ the last equation can be solved explicitly and we find 
\begin{equation}
m_d=\frac{N_fe_3^2}{2\sqrt{6}\pi}\exp[-(N_f+\frac{1}{8}+\gamma-3\log7)],
\label{32solution}
\end{equation}
and we used that
$F(1,1;2;-6)=\log7/6\,$ (note also that $\frac{1}{8} + \gamma - 3\log7
\approx -5.14$). It is obvious from comparison with Eqs.(\ref{frombelow}) and
(\ref{Aconst_mass_approx}) that our solution (\ref{32solution})
satisfies the estimates from below and above, which we obtained earlier.

Notice that up to the preexponential
factor the dependence of this solution on $N_f$ coincides with the
corresponding dependence of the (strongly coupling) solution  
(\ref{sol_N_large})
in the case of the reduced $QED_{3+1}$ with a 1-brane.
The cause of that is the fact
that, when  $N_f \gg 1$, in both cases the gauge field propagators are
dominated by the 
1-brane vacuum polarization
function, which is the same.
All the information about extra dimensions   
(like the number of
dimensions, geometry, etc.) is contained in the preexponential
factor.

The reason why the gauge field propagator is dominated by the
1-brane vacuum polarization function in the reduced $\mbox{QED}_{2+1}$
is rather subtle.
An analysis of the gap equation (\ref{A32SDeq:mass})
for the running mass function shows that its nontrivial
solution is formed by momenta on the interval
$(m_d \sqrt{6}, \Lambda)$: this equation with the low ultraviolet
cutoff $m_d \sqrt{6}$ does not have a nontrivial solution.
In the limit $N_f \gg 1$, the vacuum polarization
dominates on the interval $(m_d \sqrt{6}, \Lambda)$, and the
equation reduces to
a simple Gross-Neveu like equation whose solution
is $m_d \sim \exp{(-N_f)}$.

Notice that this is not true for the constant mass approximation,
where a nontrivial mass is generated even for the low
ultraviolet cutoff $m_d\sqrt{6}$.
Therefore, the constant mass approximation in this
case gives a different result for the dynamical mass
(\ref{Aconst_mass_approx}) than
the correct solution for the running mass function (\ref{32solution}).
This is unlike the case of the reduced $\mbox{QED}_{3+1}$ with a
1-brane, where the constant mass approximation is reliable.

\end{document}